\documentclass[conference]{IEEEtran}
\IEEEoverridecommandlockouts
\usepackage{cite}
\usepackage{amsmath,amssymb,amsfonts}
\usepackage{algorithmic}
\usepackage{graphicx}
\usepackage{textcomp}
\usepackage{xcolor}
\def\BibTeX{{\rm B\kern-.05em{\sc i\kern-.025em b}\kern-.08em
    T\kern-.1667em\lower.7ex\hbox{E}\kern-.125emX}}
\begin{document}

\title{Avatar-centred AR Collaborative Mobile Interaction
}

\author{
\IEEEauthorblockN{Bianca Marques}
\IEEEauthorblockA{\textit{DI, FCT NOVA } \\
\textit{ Universidade Nova de Lisboa }\\
Lisboa, Portugal \\
bl.marques@campus.fct.unl.pt }
\and
\IEEEauthorblockN{Rui Nóbrega}
\IEEEauthorblockA{\textit{NOVA LINCS, FCT NOVA } \\
\textit{ Universidade Nova de Lisboa }\\
Lisboa, Portugal \\
rui.nobrega@fct.unl.pt}
\and
\IEEEauthorblockN{Carmen Morgado}
\IEEEauthorblockA{\textit{NOVA LINCS, FCT NOVA } \\
\textit{ Universidade Nova de Lisboa }\\
Lisboa, Portugal \\
cpm@fct.unl.pt }
}

\maketitle

\begin{abstract}
Interaction with the physical environment and different users is essential to foster a collaborative experience. For this, we propose an interaction based on a central point represented by an Augmented Reality marker in which several users can capture the attention and interact with a virtual avatar. The interface provides different game modes, with various challenges, supporting a collaborative mobile interaction. The system fosters various group interactions with a virtual avatar and enables various tasks with playful and didactic components.\\
\\A interação com o ambiente físico e com diversos utilizadores é fulcral para fomentar uma experiência colaborativa. Para tal, propomos uma interação baseada num ponto central representado por uma marca de Realidade Aumentada em que vários utilizadores podem captar a atenção e interagir com um avatar virtual. A interface é adaptada para diferentes modos de jogo, com diversos desafios, proporcionando uma interação móvel colaborativa. O sistema tem o intuito de fomentar várias interações em grupo com um avatar virtual, podendo realizar várias tarefas com componentes lúdicas e didáticas.

\end{abstract}


\section{Introdução}
A interação e a partilha de informação é considerada uma necessidade humana e acontece no nosso dia a dia, quer através de conversas presenciais quer através de dispositivos que nos permitem aceder às redes sociais. A interação pessoa-máquina está em constante evolução, permitindo o desenvolvimento de interfaces que auxiliam o trabalho colaborativo auxiliado por computador (CSCW)\cite{b1}. 

As tarefas em grupo mais simples podem beneficiar do CSCW com o uso da tecnologia de Realidade Aumentada (RA) para proporcionar experiências mais interativas~\cite{b14} e imersivas em áreas como entretenimento, museus e eventos culturais~\cite{b2}.


Neste artigo apresentamos um sistema focado na partilha de um avatar virtual que será disputado simultaneamente por vários utilizadores. Pretende-se estudar diferentes tipos de interfaces móveis para fomentar a interação colaborativa entre os vários utilizadores. Os utilizadores são conduzidos através de uma narrativa com um objetivo muito específico, com as várias ações sincronizadas, em tempo real, nos dispositivos móveis. A história complementa a aplicação de forma a fomentar interações com o avatar virtual e entre os utilizadores. Também existe um sistema de pontos de forma a que os utilizadores fiquem imersos e que tenham uma maior ligação com o enredo. 


\section{Trabalho Relacionado}
A RA está em grande crescimento em distintas áreas de aplicação, tais como: a medicina, educação e formação, entretenimento, marketing, cultura e comércio, entre outras\cite{b3, b4, b5}. A  disponibilização de diversos kits de desenvolvimento de software como o Google, a Apple e o Vuforia permitiram uma maior facilidade na criação de aplicações em RA.

Independentemente da área de aplicação, os utilizadores esperam que a informação visualizada seja apresentada de forma natural, instintiva e agradável \cite{b2}. 

\paragraph{Realidade Aumentada em Dispositivos Móveis}
Um sistema MAR (\textit{Mobile Augmented Reality}) apresenta algumas limitações importantes a salientar pois apesar da portabilidade, existem fatores como a bateria, a capacidade de renderização e conetividade à internet\cite{b4, b5} que impedem o aproveitamento da experiência pelo utilizador. 

Existem cada vez mais sistemas MAR, um dos exemplos de sucesso entre o público geral é o Pokémon Go. Outro exemplo MAR de uma área diferente em RA que inspirou a criação do sistema foi \noindent o MagicBook\cite{b6} que usa um livro real para transportar o utilizador entre a realidade e virtualidade. Este não utiliza um dispositivo móvel mas utiliza óculos RA. \noindent O BBC Civilisations AR\cite{b7} engloba peças de arte de vários países, sendo uma plataforma genérica, não estando associada a um museu específico. \noindent O The Speaking Celt\cite{b8} é baseado em avatares que guiam os visitantes de um museu, apresentando informações acerca dos artefactos e da sua história.

\paragraph{Realidade Aumentada Colaborativa}
Segundo Renevier et al.\cite{b9} uma colaboração eficiente requer que cada colaborador tenha um ponto de vista único para os dados apresentados. Quando há a combinação da colaboração com RA, os colaboradores preferem partilhar o mesmo espaço físico. Existem 2 tipos de colaboração em RA, a presencial e a remota. A colaboração presencial é muito focada na interação verbal ou corporal entre vários utilizadores presentes num mesmo espaço físico. Por sua vez, na colaboração remota, não existe partilha de espaço físico e torna-se difícil a interpretação de certos gestos, podendo haver perda de informação entre os utilizadores. Um exemplo de colaboração presencial é a sala Colab na Xerox\cite{b10}. O  projeto de Studierstube\cite{b11} é um dos primeiros sistemas de RA colaborativos, sendo um exemplo de colaboração em RA presencial.

\section{Sistema Colaborativo Móvel em RA}
A solução é centrada no uso de marcas para a criação de um sistema que permite vários utilizadores partilharem simultaneamente um avatar virtual, no mesmo espaço físico. Todas as ações realizadas pelo utilizador são sincronizadas pela rede, em tempo real.
A interface tem 2 tipos de ações, curtas, como por exemplo dar toques no ecrã e, longas, como jogar um mini jogo. As últimas são acessíveis através de um menu de botões disponível na interface.

O sistema apresenta diferentes modos de jogo com diversos objetivos e interações específicas para ajudar a ter diferentes perspetivas na avaliação. Um dos modos de jogo é focado numa narrativa e outro é focado na captação de atenção do avatar. A narrativa proporciona um ambiente que o utilizador consegue controlar, criando uma ligação com o enredo tornando-o mais imersivo. O fluxo da história permite que o utilizador tenha uma missão para cumprir ao executar várias ações e ir ganhando pontos. A missão é focada em 20 pontos colaborativos em ações curtas e/ou longas para encontrar um artefacto perdido. 

O sistema foi construído usando Unity com componentes adicionais que permitiram o uso da tecnologia de RA, neste caso o Vuforia. Este SDK permite guardar as marcas na própria base de dados onde são feitas as extrações das \textit{features} para permitir a detetação da imagem. Quanto maior o número de \textit{features} mais fácil é a detetação da imagem. Também foi utilizado, o motor de rede Photon PUN para sincronizar, em tempo real, as ações realizadas por cada utilizador. 
Para criar novas formas de interação foi utilizado um \textit{plugin} para o reconhecimento de voz, o \textit{Speech And Text Unity iOS Android}
. O sistema permite configurar, de forma simples, a história e toda a narrativa e enredo, bastando para tal modificar um ficheiro JSON, alterando completamente o contexto do sistema, tornando-o numa experiência nova. A arquitetura do sistema é visível na Figura 1.

\begin{figure}[t]
\centerline{\includegraphics[width=1\linewidth]{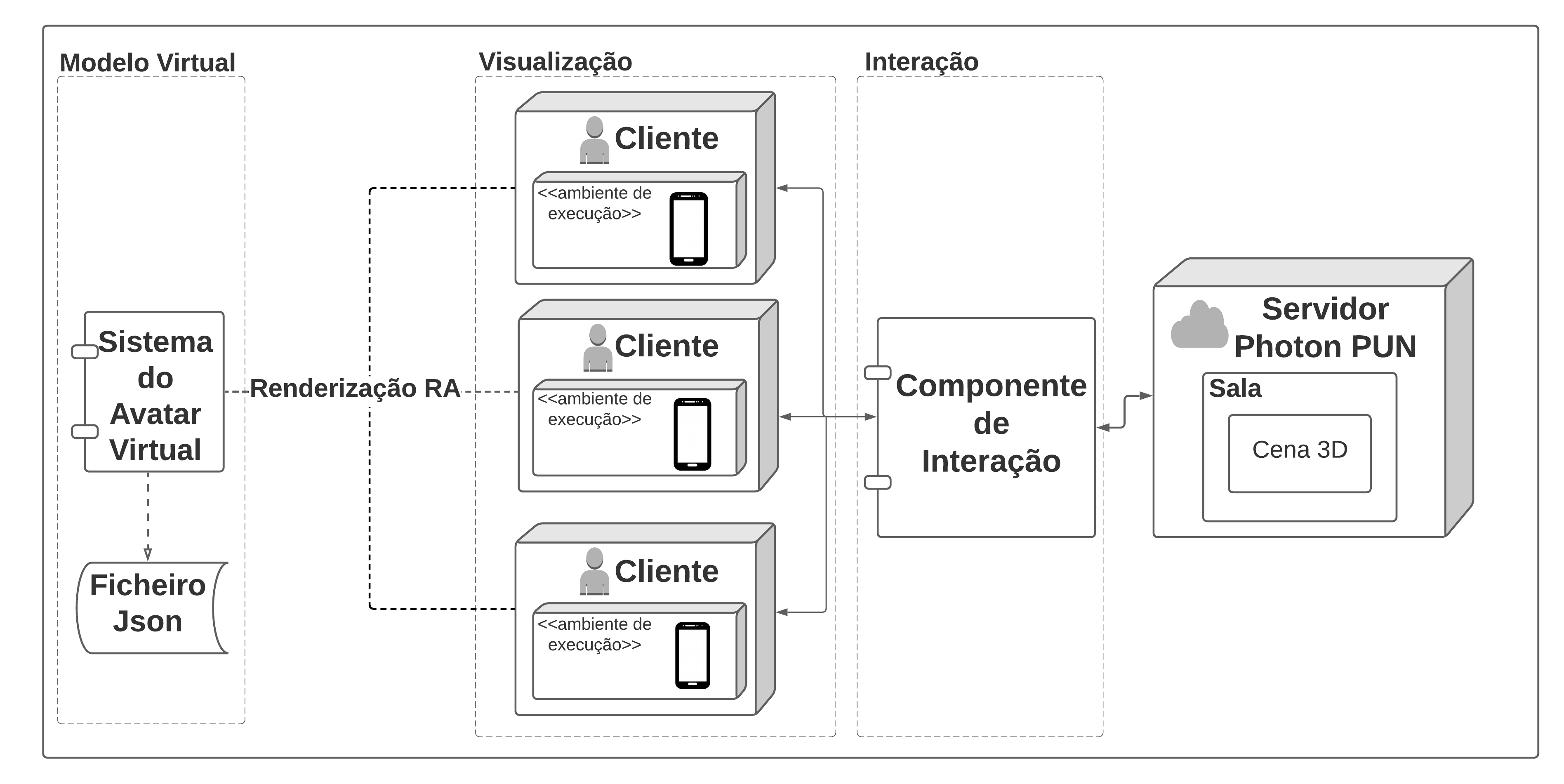}}
	\caption{Arquitetura do sistema composta pela marca de RA, pelos clientes, pelas interações de cada e pelo servidor que sincroniza essas interações.}
\label{fig: arquitetura}
\end{figure}

\section{Design de Interface e Funcionalidades}
Um dispositivo móvel deve ter uma interface intuitiva de forma a que o utilizador tenha uma experiência facilitada diminuindo o tempo de aprendizagem relativa ao sistema. O design da interface foi pensado no conforto e estabilidade do dispositivo móvel nas mãos do utilizador, por isso, a orientação padrão é horizontal. Assim, o utilizador possuiu uma área maior de jogo e de visualização.

Observando a Figura 2, as ações longas (minijogos) são representadas por botões amarelos no lado direito, enquanto as ações curtas (toques e deslizes no ecrã) são representadas por toda a área da tela excluindo o menu. Este método de divisão foi pensado na usabilidade e facilidade do utilizador ao interagir com a interface. 

Quanto às funcionalidades de ações curtas, estas são:
\begin{itemize}
    \item 1 toque - o avatar dança ao som da "Macarena".
    \item 2 toques - o avatar dança ao som do "Samba".
    \item 3 toques - o avatar dança ao som do "I like to move it".
    \item Deslizar no ecrã em qualquer direção - o avatar dança ao som do "Twist".
    \item 4 toques ou mais - o utilizador lança um pequeno caos, a animação/dança é escolhida de forma aleatória com o conjunto dos pontos acima.
    \item 4 toques ou mais -  o utilizador lança o caos, o avatar dança ao som do "Axel F" do \textit{Crazy Frog}.
\end{itemize}

Quanto à funcionalidade de 4 ou mais toques, a decisão para ser pequeno caos ou caos é calculada consoante as ações do utilizador até ao momento, por outras palavras, é a divisão do número de toques individuais até ao momento pelo número total de utilizadores na sala virtual. 
A ação curta caos tem como objetivo criar um elemento de aleatoriedade e choque no jogo. Esta ação depende de condições específicas no jogo. Haverá uma maior probabilidade de um utilizador visualizar o caos, se houver um grande número de utilizadores ou um número reduzido de ações realizadas individualmente.

As ações curtas podem ser exploradas em todos os modos, mas são exclusivas para o Modo Toques e para o Modo Avatar. As ações longas são minijogos específicos do Modo Surpresa e são descritas na subsecção seguinte do Modo História.

Sempre que um utilizador entra numa sala virtual é-lhe atribuída uma cor única. A cor é o identificador de cada utilizador e surge no canto inferior esquerdo da interface, como pode ser observado na Figura 2. Sempre que um utilizador realiza uma ação surge uma notificação com a cor do jogador que realizou a ação e a ação realizada, como mostrado no exemplo na Figura 3. 

\begin{figure}[t]
\centerline{\includegraphics[width=1\linewidth]{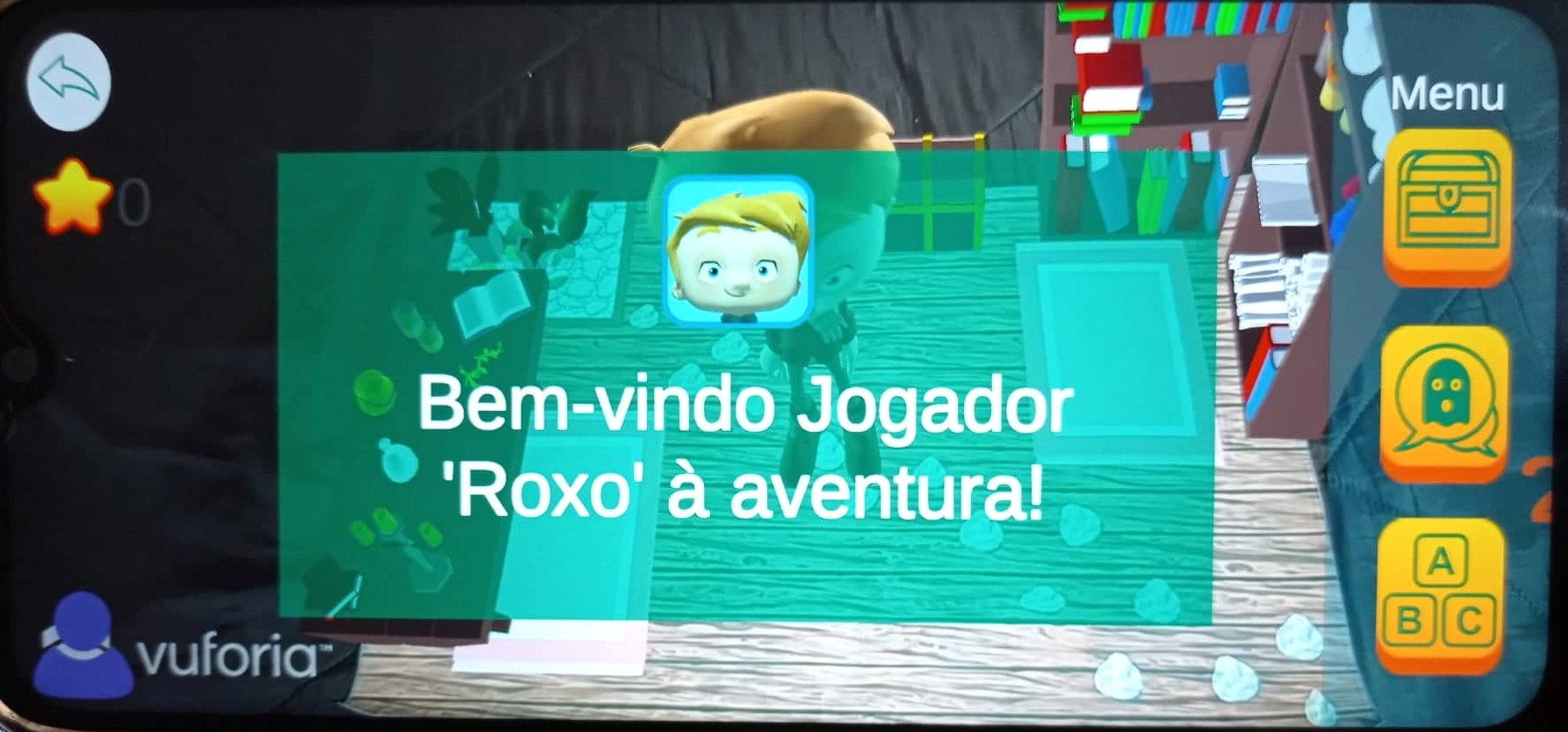}}
	\caption{Mensagem inicial quando um utilizador entra numa sala virtual e é-lhe atribuída uma cor identificadora.}
\label{fig: arquitetura}
\end{figure}

\begin{figure}[t]
\centerline{\includegraphics[width=1\linewidth]{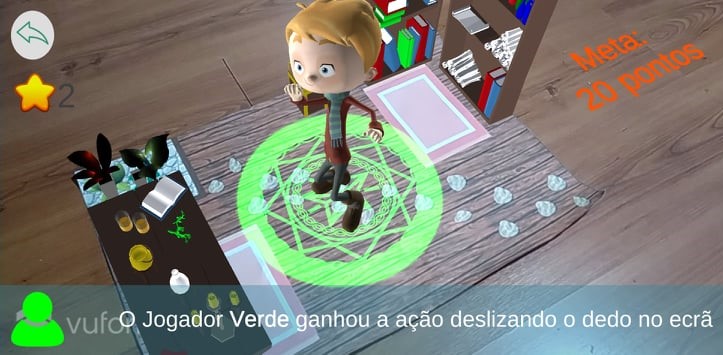}}
	\caption{Notificação de ação. Avatar virtual virado em direção do utilizador que executou a ação.}
\label{fig: arquitetura}
\end{figure}

Relativamente a melhorar a experiência e imersão do utilizador, para além da interface, foram analisadas as dimensões e o design das marcas a utilizar. O design foi pensado tendo em conta a narrativa e uma imagem que não tenha um formato simples, nem seja simétrico ou aborrecida como um QRCode. O utilizador pode interagir com duas marcas, a marca principal e marca que é usada num dos minijogos. Quanto à marca principal, impressa em A2, proporciona uma maior dimensão do mundo virtual, tornando-o o mais realista possível para o utilizador. Quanto à outra marca, está disponível para o utilizador pegar e mover, sendo por isso de menor dimensão. 

Quando o utilizador inicializa a aplicação, tem duas opções de escolha, o Modo Toques e o Modo História. Dentro da História tem o Modo Avatar e Modo Surpresa.

\subsection{Modo Toques}
Quando o utilizador clica no Modo Toques não tem qualquer informação sobre como interagir com o avatar, ou seja, a chave neste modo é a descoberta das possíveis ações curtas que permitem interagir com o avatar. Durante as interações em grupo, cada utilizador compete pela atenção do avatar enquanto executa as ações curtas descobertas.

\subsection{Modo História}
Quando o utilizador clica no Modo História é-lhe apresentado a narrativa com a apresentação do desafio: interagir com o avatar para perfazer 20 pontos. Neste modo, quase todas as ações curtas valem 1 ponto, exceto a ação curta que lança o pequeno caos que vale 2 pontos e a ação curta que lança o caos que vale 3 pontos. Uma leve explicação do objetivo e de acções curtas e longas também é apresentada num pequeno tutorial. Após o tutorial, o utilizador pode escolher entre o Modo Avatar e o Modo Surpresa, de notar que o desafio é comum a ambos os modos.

\subsubsection{Modo Avatar}
No Modo Avatar o utilizador interage com o avatar virtual enquanto tenta atingir a missão proposta na narrativa. Esta pode ser alcançada rapidamente caso note que certas ações curtas lhe permite ganhar mais pontos.

\subsubsection{Modo Surpresa}
No Modo Surpresa, o utilizador tem a possibilidade de realizar ações curtas e/ou ações longas. Nas ações longas, são apresentadas caixas de texto para explicar o minijogo, estas guiam o utilizador. É de notar que caso o utilizador escolha um minijogo, a possibilidade de ganhar mais pontos é maior, podendo ganhar até 4 pontos. 
As ações longas podem ser sequenciais ou aleatórias. Também podem ser combinadas, onde uma parte é sequencial e outra é aleatória, consoante as várias etapas para completar uma ação longa.

Quanto às funcionalidades de ações longas, estas são:

\paragraph{Minijogo Objetos Escondidos} Ação Combinada. O desafio é encontrar a chave para abrir o baú do cenário virtual. Para o realizarem necessitam de interagir com segunda marca, a imagem da avó, e posicioná-la corretamente e encontrar todos os objetos escondidos.

\paragraph{Minijogo Fantasminha} Ação Sequencial. O desafio é responder corretamente ao máximo de perguntas usando o microfone. Cada utilizador responde às perguntas no seu dispositivo contudo podem interagir e ajudar-se pois no final o ganho de pontos é colaborativo. 

\paragraph{MiniJogo Descobrir a Palavra} Ação Aleatória. Este minijogo é inspirado no jogo da forca. O desafio é descobrir a palavra sabendo o número de letras, contudo apenas têm 7 tentativas para a descobrir. 

Algo comum em todos os modos é a capacidade de ser visível qual foi o utilizador que realizou a ação através da notificação e da rotação do avatar em direção à pessoa que executou a ação, como observado na Figura 3. 

\section{Avaliação}
Com o objetivo de analisar a usabilidade do sistema e a influência do trabalho em equipa foram realizados 2 tipos de testes, um singular e o outro em grupo de duas pessoas sobre uma população de 26 pessoas (13 grupos).

Analisando a caraterização da população total, podemos aferir que 54\% era do género masculino enquanto o restante era do género feminino, com mediana de idades de 22 anos. Metade dos utilizadores nunca tinham tido contato com um sistema em Realidade Aumentada e os que tiveram contacto nomearam sempre o Pokémon Go. 

No teste singular foi utilizado o Modo de Toques (Ver subsecção IV.A). O gráfico da Figura 4 mostra as ações curtas não detetadas e a avaliação do utilizador à pergunta da legenda. Observando o gráfico, podemos destacar que 57.7\% da população discordou totalmente com a afirmação enquanto 7.7\% da população nem chegou a descobrir a ação. Outro dado a destacar é a dificuldade em descobrir a ação curta caos, 34.6\% da população. Neste modo, foram recolhidas as expressões corporais e avaliadas de 1 a 5 quanto à positividade (movimentos alegres, dançar ou rir), com mediana 4.
O teste em grupo no Modo Toques, permitiu aferir que 8 dos 13 grupos (61.5\%) realizaram as ações curtas todas enquanto disputavam pela atenção do avatar. As ações curtas não realizadas foram 2 e 3 toques com igual percentagem. Neste modo, foram recolhidas as expressões corporais individuais e avaliadas de 1 a 5 quanto à positividade, a mediana foi de 5.

No teste em grupo no Modo Surpresa (Ver subsubsecção IV.B.2), o modo foi eleito como o preferido por todos os utilizadores, com 50\% da população a classificar o minijogo dos objetos escondidos como o melhor. O segundo melhor, com 38.5\%, o minijogo de descobrir as palavras. E por último, o minijogo do fantasminha apenas com 11.5\% contudo todos os utilizadores acharam a interação colaborativa positiva. Em todos os minijogos a interação foi positiva, avaliada com uma média maior ou igual a 4.7.

Comparando os testes colaborativos, no Modo Toques os utilizadores disputavam pela atenção do avatar e todas as componentes como a rotação do avatar e a notificação foram importantes para os resultados positivos enquanto que no Modo Surpresa, os utilizadores comunicaram muito mais e ajudaram-se mutuamente a realizar as ações mais complexas. Os resultados mostram que os utilizadores preferem realizar ações que tenham um impacto mais visual e que as ações mais básicas como 1 toque ou deslizar o dedo são intuitivas para o utilizador explorar o sistema sem precisar de ajuda. Isto permitiu obter no questionário de usabilidade SUS \cite{b13}, uma pontuação de 85.87. Interpretando esta pontuação, segundo Bangor et al.\cite{b15}, o sistema é considerado "Aceitável" na escala de aceitabilidade, "Excelente" na classificações de adjetivos e de nota B.

\begin{figure}[t]
\centerline{\includegraphics[width=1\linewidth]{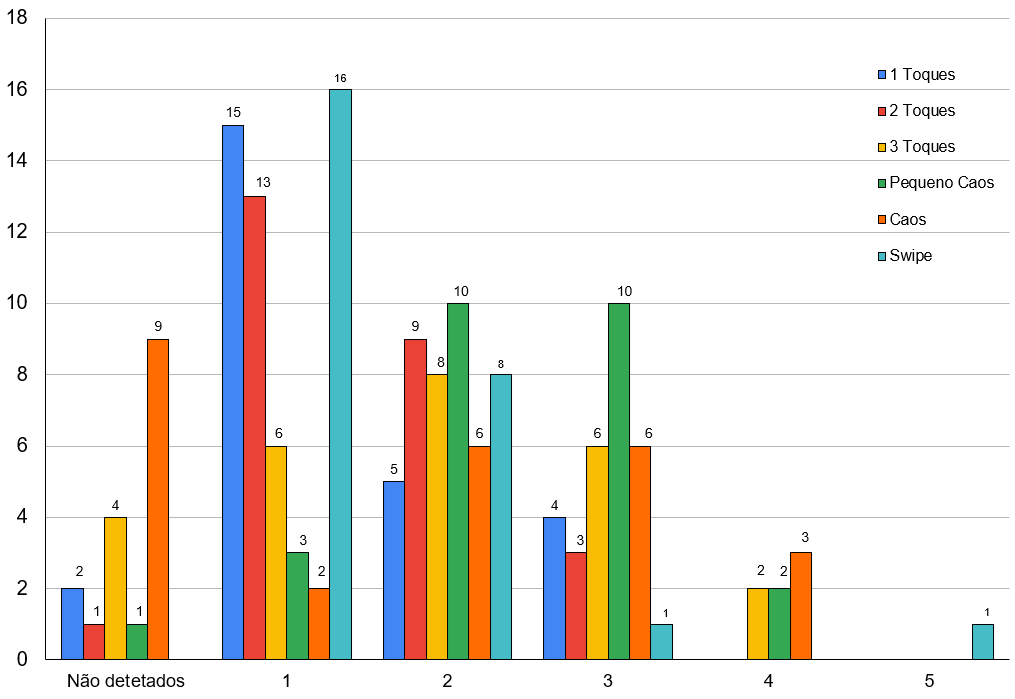}}
	\caption{Gráfico relação avaliação de dificuldade ação curta/cada ação curta Modo Toques. Avaliação de 1 (discordo totalmente) a 5 (concordo totalmente) à afirmação "A funcionalidade X foi difícil de executar."}
\label{fig: chart}
\end{figure}

\section{Conclusão e Trabalho Futuro}
Os resultados dos testes dos utilizadores mostraram a positividade na interação em diferentes modos de jogo, revelando algumas interações mais benéficas que outras, nomeadamente as ações mais simples, foram as mais intuitivas e as ações mais complexas foram as mais apelativas para os utilizadores. Isto só foi possível devido às informações visuais e auditivas terem fomentado o impacto positivo da RA (Ver Tabela \ref{tab1}). 

Seria interessante explorar a interação com objetos físicos do mundo do real ou usar mais marcas de RA para fomentar a componente didáctica no sistema ou alterar a interação do microfone. No geral, o sistema atual pode ser utilizado por uma ou várias pessoas, sem limite de idade e sem necessidade de aprendizagem acerca do sistema. Isto mostra que apesar de alguns utilizadores não conhecerem nem estarem habituados a aplicações em RA, esta é uma boa aposta para implementar diversas experiências, inovando-as e beneficiando com elas.

\begin{table}[t]
\caption{Resultados da interação em grupo}
\label{tab: tab1}
\begin{center}
\begin{tabular}{|p{0.6\linewidth}|c|c|c|}
\hline
\textbf{Pergunta}&\textbf{Q1} &\textbf{Mediana} &\textbf{Q3} \\
\hline
\textbf{"O sistema ajustou-se corretamente ao dispositivo que utilizei."} & 4 & 4 & 5\\
\hline
\textbf{"Os aspetos visuais envolveram-me no sistema."} & 4 & 5 & 5\\
\hline
\textbf{"Os aspetos sonoros envolveram-me no sistema."} & 4 & 5 & 5\\
\hline
\textbf{"Considero que no geral, a interação colaborativa neste sistema foi positiva."} & 5 & 5 & 5\\
\hline
\end{tabular}
\label{tab1}
\end{center}
\end{table}

\section*{Agradecimentos}
Gostaríamos de agradecer o apoio do centro de investigação NOVA LINCS com o projeto UIDB/04516/2020.

\vspace{12pt}

\end{document}